\documentclass[seceq]{ptptex}

\def\etal{{\it et al.}\rm}

\def\M{m_{\mbox{\scriptsize{Pl}}}}
\def\ltsima{$\; \buildrel < \over \sim \;$}
\def\gtsima{$\; \buildrel > \over \sim \;$}
\def\simlt{\lower.5ex\hbox{\ltsima}}
\def\simgt{\lower.5ex\hbox{\gtsima}}
\def\pmb#1{\setbox0=\hbox{#1}%
 \kern-0.05em\copy0\kern-\wd0
 \kern0.20em\copy0\kern-\wd0\box0}
\def\KKLMMT{\pmb{K}L\pmb{M}T$\;\;$}

\title{The Search For Primordial Tensor Modes}

\author{George Efstathiou and Sirichai Chongchitnan}

\inst{Institute of Astronomy, Madingley Road, Cambridge, CB3 OHA. \\ England.}

\abst{We review the prospects for detecting tensor modes generated
during inflation by CMB polarization experiments and by searching for
a stochastic gravitational wave background with laser interferometers
in space. We tackle the following two questions: (i) what does inflation
predict for the tensor fluctuations? (ii) is it really worth building experiments
that can cover only a small range of tensor amplitudes?}

\begin{document}

\maketitle

\section{Introduction}

Inflation is an extremely attractive idea that has gained widespread
support.
However, even a cursory glance at the literature will reveal a
plethora of inflationary models. Inflation theory is mired
in phenomenological model building (often involving `unnatural'
fine-tunings) rather than emerging in a compelling way from
fundamental physics.

From the observational point of view, inflation has received strong
support from observations of the cosmic microwave background (CMB)
anisotropies, and from studies of the large-scale distribution of
galaxies. In particular, the beautiful CMB results from WMAP
\cite{spergel} are consistent with primordial adiabatic fluctuations
with a nearly scale invariant spectrum, as expected in the simplest
inflationary models. A key prediction of inflationary models is the
existence of a stochastic background of gravitational waves. Such a
background has not yet been observed, but its detection would provide
incontrovertible evidence that inflation actually occurred and would
set strong constraints on the dynamics of inflation. It is therefore
no surprise that a vigorous effort is underway to detect tensor modes
from inflation.

In this article, we will first review what can be learned about
inflationary models from the detection of tensor modes. We will then
discuss the prospects for detecting tensor modes from observations of
the polarization of the CMB and by direct detection using
interferometers in space. We will then tackle the following thorny
questions:

\medskip

\noindent
(i) What does inflation predict for the amplitude of the tensor fluctuations?

\medskip

\noindent
(ii) If theory does not constrain the amplitude of the tensor mode
to within many orders of magnitude, is it really worth the effort to build 
experiments that can only cover a small range?

\medskip

Our perspective on point (i) is very different to that presented in a
recent {paper\cite{boyle1}}.
Point (ii) is
clearly important in assessing the case for a post-Planck satellite
dedicated to CMB polarization {measurements\cite{petersen}}.  Unless otherwise
stated, we will assume the `concordance' $\Lambda$-dominated cold dark
matter cosmology, with cosmic densities as given {in\cite{spergel}}.

\section{Gravitational Waves from Inflation}

For the most part, we will consider only the simplest single field
inflationary models, characterised by a potential $V(\phi)$ and
Hubble constant $H(\phi)$, where $\phi$ is the inflaton
field. Following  the normalizations
of{\cite{lidsey}\cite{stewart}}, the amplitudes $A(k)$ of scalar and
tensor power spectra generated during inflation are given, to lowest
order, by
\begin{eqnarray}
 A_S(k) &\approx&  {4\over5}{H^2\over \M^2|H'|}\bigg|_{k=aH} ,\\
 A_T(k) &\approx&  {2\over5\sqrt{\pi}}{H\over \M}\bigg|_{k=aH},\label{at}
\end{eqnarray}
where $S$ and $T$ denote scalar and tensor components respectively,
 $\M$ is the Planck mass, and primes denote derivatives with respect
 to  $\phi$.  The amplitudes in (\ref{at}) are
 evaluated when each mode $k$ is equal in scale to the Hubble radius,
 {\it i.e.} when $k=aH$.  The spectral indices $n_s$ and $n_T$ are
 defined by
\begin{eqnarray}
 n_s -1 &\equiv& {d\ln A_S^2(k)\over d\ln k},\\
n_T &\equiv& {d\ln A_T^2(k)\over d\ln k} . \label{ind1}
\end{eqnarray}
evaluated at a `pivot' scale $k_0$ (which we will take to be $k_0 =
0.002\; {\rm Mpc}^{-1}$, {following\cite{peiris}}). To first order in the `slow
roll' parameters,
\begin{eqnarray}
 \epsilon \equiv {\M^2\over 4\pi}
\left({H'\over H}\right)^2
, \qquad\eta \equiv {\M^2\over 4\pi}
\left({H''\over H}\right), \label{ind2}
\end{eqnarray}
one finds that the spectral indices are given by
\begin{equation} 
 n_s -1 \approx 2\eta-4\epsilon, \quad n_T \approx -2\epsilon.
\end{equation}
Following the convention {of\cite{peiris}}, we define the tensor-scalar
ratio $r$, as
\begin{equation}
r = 16 {A_T^2\over A_S^2}\approx 16 \epsilon, \label{tens1}
\end{equation} 
where the last relation applies to first order in `slow-roll'
parameters.  Note that the definition of a `tensor-scalar' ratio
varies widely in the literature. For instance, it is often
defined\cite{liddle} as the ratio of the tensor and scalar CMB
quadrupoles $r_2=C^T_2/C^S_2$. Such a definition is
dependent on cosmology, especially on the dark energy density
$\Omega_\Lambda$. (See\cite{turnerwhite} for a relation between $r_2$
and $r$.)

As is well-known, the Thomson scattering of an anisotropic photon
distribution leads to a small net linear polarization of the CMB
anisotropies. (See\cite{hu} for an introductory review and 
references to earlier work.) This polarization signal can be decomposed
into scalar $E$-modes and pseudo-scalar $B$-modes. The separation of a
polarization pattern into $E$ and $B$ modes is of particular interest
since scalar primordial perturbations generate only $E$ modes while
tensor perturbations generate $E$ and $B$ modes of roughly comparable
amplitudes{\cite{zaldarriaga}\cite{kamionkowski}}.

\begin{figure}

\vskip 4.01 truein

\includegraphics{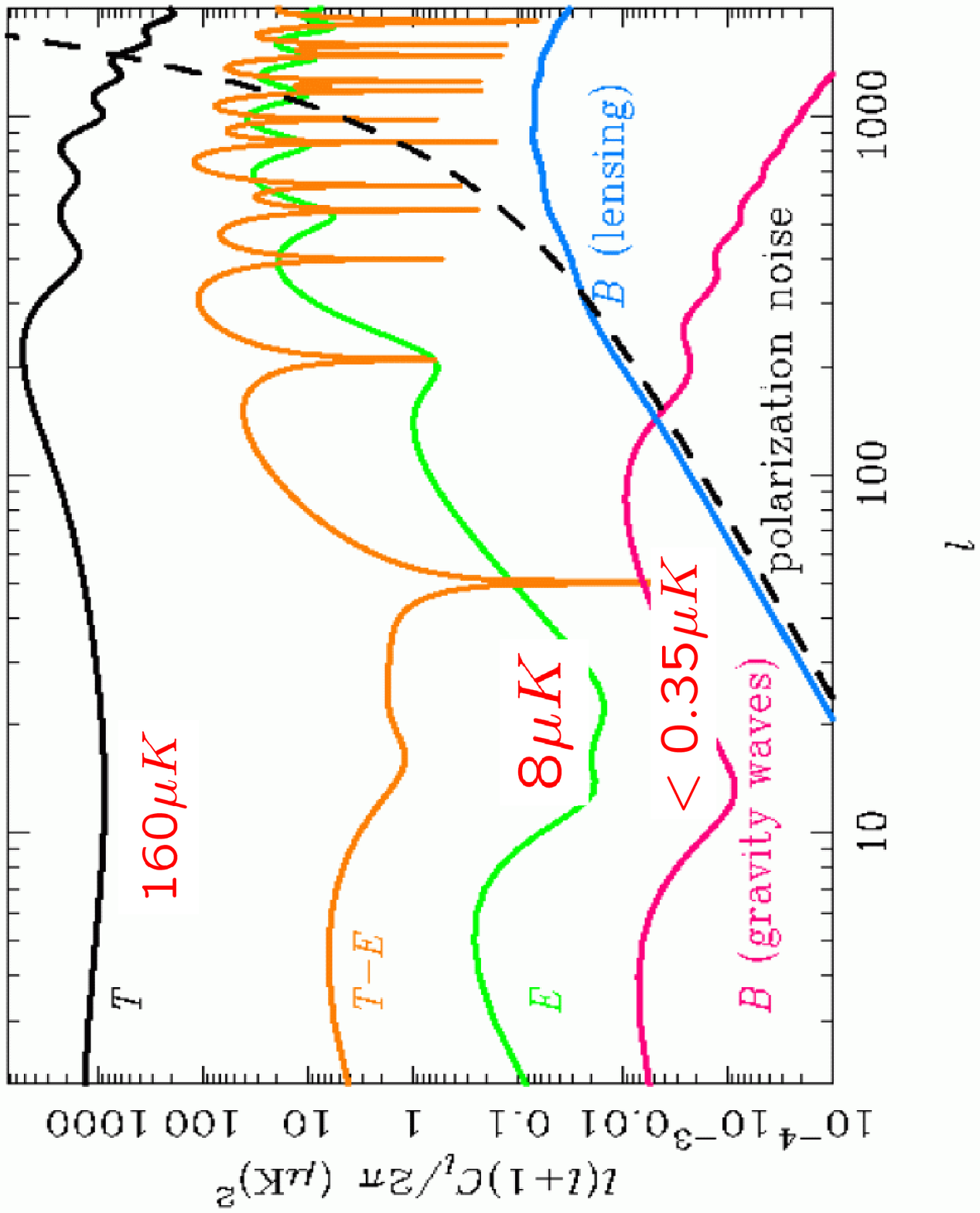}

\caption{Temperature and polarization power spectra for the
concordance $\Lambda$CDM model.  The current (indirect) upper limit of
\cite{seljak} leads to an upper limit on {\it rms} polarization
anisotropy of $\simlt 0.35 \mu{\rm K}$. For comparison, the {\it rms}
signals for the $T$ and $E$ anisotropies are given. Lensing of $E$
modes by intervening matter leads to small scale $B$
modes\cite{zaldarriaga2} as shown. The dashed line the white-noise
level for an experiment with a sensitivity to $B$-modes of $r \sim
10^{-2}$.}
\label{Figure1}
\end{figure}

\smallskip

The $T$, $TE$, $E$, CMB power spectra for the concordance $\Lambda$CDM
cosmology are shown in Figure \ref{Figure1}.  An $E$-mode polarization
signal was first discovered by DASI\cite{kovac}\cite{leitch} Exquisite
measurements of the temperature-$E$-mode cross power spectrum have
been reported by the WMAP {team\cite{kogut}}. Measurements of the
$E$-mode power spectrum have been reported by the CBI
experiment\cite{readhead} and by the 2003 flight of
{Boomerang\cite{montroy}}. Primordial $B$-mode anisotropies have not
yet been detected in the CMB. The best current upper limits come from
model fitting to CMB data and observations of the matter power
spectrum at low {redshift\cite{seljak}}, leading to the 95\% upper limit
of
\begin{equation}
 r \simlt 0.36. \label{niceobs1}
\end{equation}
A primordial $B$-mode of this amplitude would produce an {\it rms}
anisotropy signal of only $\sim 0.35 \mu{\rm K}$, {\it i.e.} about a factor
of $20$ times smaller than the {\it rms} anisotropy in $E$ modes (Figure
\ref{Figure1}). Evidently, the detection of primordial tensor modes 
presents a formidable experimental challenge. 

In the next Section, we will review briefly the prospects for 
experimentally detecting primordial tensor modes. We will then turn
to  theoretical implications and address the two questions posed
in the Introduction.

\section{Prospects for Detecting Tensor Modes}

\subsection{WMAP and Planck}

As mentioned above, the first year of data from WMAP have been used to
measure the $TE$ power {spectrum\cite{kogut}}.  Direct detection of
electric polarization, via its power spectrum $C_\ell^E$, is more
challenging than statistical detection using the cross-correlation
with the temperature anisotropies, since the expected $E$-polarization
signal is much weaker than the correlated part of the temperature
(Figure \ref{Figure1}). At the time of writing, results for $C_\ell^E$
from the first three years of data from WMAP are eagerly awaited.

\begin{figure}
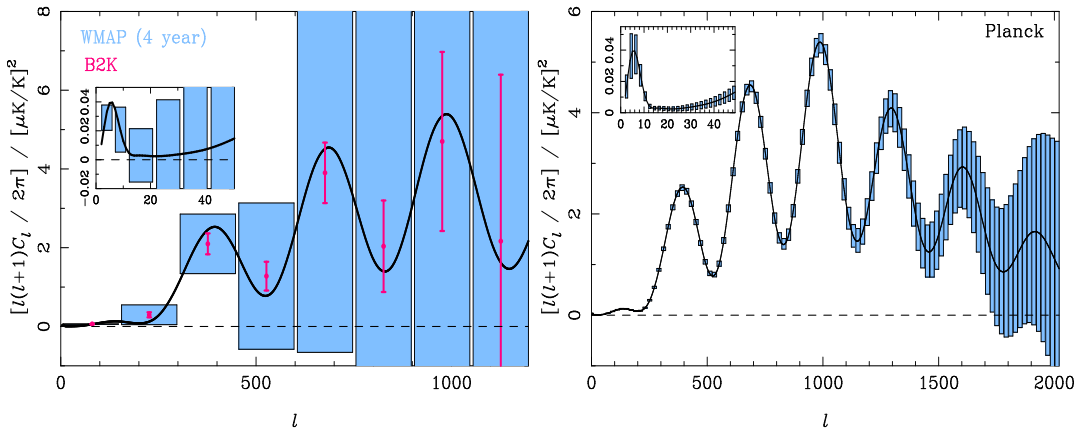


\vskip 2.5 truein

\includegraphics{figure2a.ps}
\includegraphics{figure2b.ps}

\caption{The left hand panel shows forecasts for the $\pm 1\sigma$
errors on the electric polarization power spectrum $C_l^E$ from WMAP
after 4 years of observation.  (Forecasts for the Boomerang (2003)
experiment, labelled B2K, are also plotted). The right hand panel
shows forecasts for Planck.  For WMAP and B2K, flat band powers are
estimated with $\Delta l = 150$ (with finer resolution on large scales
for WMAP in the inset). For Planck, flat band powers are estimated
with $\Delta \ell =20$ in the main plot and with $\Delta \ell=2$ in
the inset. (Figures computed by A, Challinor, reproduced from
\cite{bluebook}).}

\label{Figure2}
\end{figure}

Figure \ref{Figure2} shows {forecasts\cite{bluebook}}, using the WMAP
instrument sensitivities and beam widths, for what we might expect in
an idealised case in which polarized emission from the Galaxy can be
neglected over 65\% of the sky. The concordance $\Lambda$CDM model has
been assumed with a high optical depth for secondary reionization of
$\tau=0.17$.  After four years, WMAP may make a detection in a few
broad bands around $l=400$, and on the largest scales where
reionization dominates. Evidently, WMAP barely has the sensitivity to
measure the $E$-polarization signal let alone the much smaller signal
expected for primordial tensor modes.

The Planck {satellite\cite{bluebook}}, scheduled for launch in August
2007, has higher sensitivity to polarization as shown by the right
hand panel. Planck should be able to map out $C_l^E$ on all scales up
to and beyond the global maximum at $l \sim 1000$. However, even at the
sensitivities expected for Planck, the detection of $B$-modes will
pose a formidable challenge.

\begin{figure}

\vskip 2.9 truein

\includegraphics{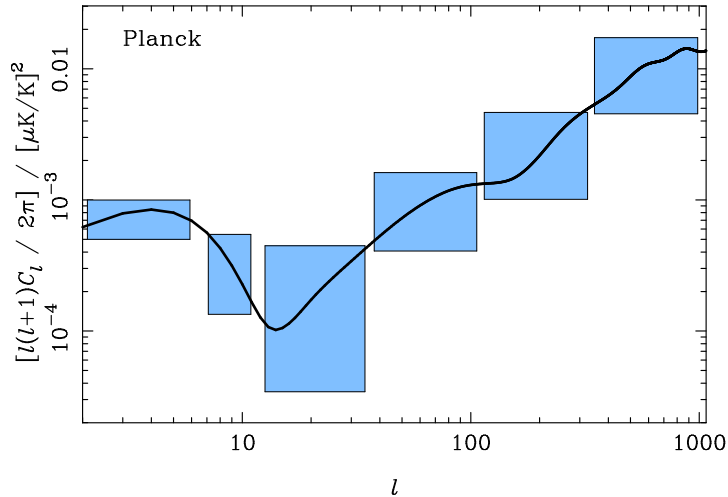}

\caption{Forecasts for the $\pm 1\sigma$ errors on the magnetic polarization
power spectrum $C_l^B$ from Planck. Above $l \sim 150$ the primary spectrum
is swamped by weak gravitational lensing of the $E$-modes\cite{zaldarriaga2}
(Figure computed by A. Challinor, reproduced from \cite{bluebook}).}
\label{Figure3}
\end{figure}

  This is illustrated by Figure \ref{Figure3}, which shows the errors
on, $C_l^B$, expected from Planck for a model with $r$ arbitrarily set
to $0.1$. Note that with such early reionization almost 50 per cent of
the total power in primordial $B$-polarization is generated at
reionization, so confirmation of the high optical depth suggested by
WMAP\cite{kogut} will have important implications for Planck.  The
figure suggests that for $r=0.1$ Planck can characterise the
primordial power spectrum in around four bands.  The $B$-polarization signal
generated by weak gravitational lensing\cite{zaldarriaga2} (which is,
of course, independent of $r$) dominates the primary signal above $l
\sim 150$. At the very least, the weak lensing signal should be
detectable by Planck. However, even if systematic errors and polarised
Galactic emission can be kept under control, Planck will at best only be able
to detect tensor modes from inflation if the tensor-scalar ratio is
greater than a few percent.

\subsection{More Sensitivity?}

In case the reader finds the discussion above a little depressing, it
is important to stress that constraining $r$ to within a few percent
via CMB polarization would be a considerable
achievement. Nevertheless, one can ask whether it is possible to do
better.  The key to achieving even higher sensitivities than Planck is
to build experiments with large {\it arrays} of polarization sensitive
detectors. Two such experiments, both ground-based, are known to the
authors. Clover\cite{taylor} (the `Cl-Observer') which will use large
bolometer arrays and QUIET\footnote{QU Imaging ExperimenT, 
http://quiet.uchicago.edu.} which
will use large arrays of coherent detectors.

\begin{figure}

\vskip 3.3 truein

\includegraphics{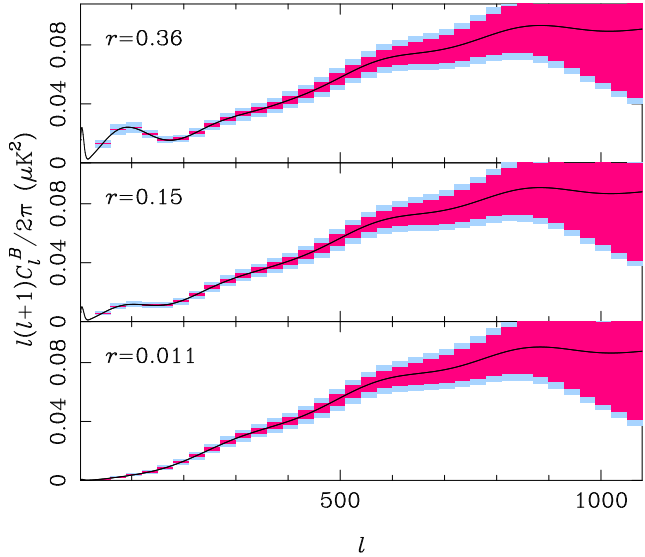}

\caption{ The expected errors from Clover on the $B$-mode power
spectrum. The upper panel has tensor-scalar ratio $r=0.36$ ({\it
cf.} equation \ref{niceobs1}), the middle panel is for $r=0.15$ and
the lower panel is for $r=0.011$.  The smaller (magenta) error boxes
are the contribution from instrument noise and the larger (blue) boxes
also include sample variance, include the contribution from weak
lensing. (Figure computed by A. Challinor).}
\label{Figure4}
\end{figure}

Both of these experiments are still under development, and so there
are uncertainties in forecasting what they might see. At present, it
is envisaged that Clover will operate at three frequencies, $97$,
$150$ and $230$ GHz, with a resolution of $10$ arcminutes. Each focal
plane will be populated by a hexagonal array of corrugated single-mode
feed horns, $160$ at $97$ GHz, $260$ at each of $150$ GHz and $230$
GHz, whose outputs will be detected using novel arrays of
voltage-biased Transition Edge Sensors.  Figure \ref{Figure4} shows
forecasts computed by Anthony Challinor, for various values of the
gravitational wave amplitude, assuming one-year of integration with
the full instrument and including the effects of foreground
subtraction. This shows that Clover should have sufficient thermal
sensitivity that its measurement of the $B$-mode power spectrum should
be limited by lensing variance up to $l=200$, covering the range where
gravity waves contribute. These calculations suggest that
a tensor-scalar ratio of $\sim 0.01$
should be detectable with this instrument at about the $3\sigma$
level. There are, therefore, good prospects for achieving high
precision limits on primordial $B$-modes from future ground-based
experiments.

\subsection{Direct Detection with Interferometers in Space}

Another way of detecting tensor modes generated during inflation is to
search for a stochastic gravitational-wave background using laser
interferometers (see, for example, the review by Cooray\cite{cooray}
and references therein; for recent discussions
see\cite{smith1}\cite{boyle2}\cite{smith2}). The expected
gravitational wave spectrum for wavenumbers  
$k \gg k_{\rm equ}$ {is\cite{chongchitnan}},
\begin{equation}
 \Omega_{\rm gw}(k) \approx {375\over 4H_0^2 k^2_{\rm equ}
\tau_0^4}\langle \vert A^2_T(k) \vert^2 \rangle, \label{gw1}
\end{equation}
where $\tau_0$ is the conformal time at the present day and $k_{\rm
equ} = \tau_{\rm equ}^{-1}$ is the wavenumber that equals the Hubble
radius at the time that matter and radiation have equal densities.

\begin{figure}

\vskip 3.5 truein

\includegraphics{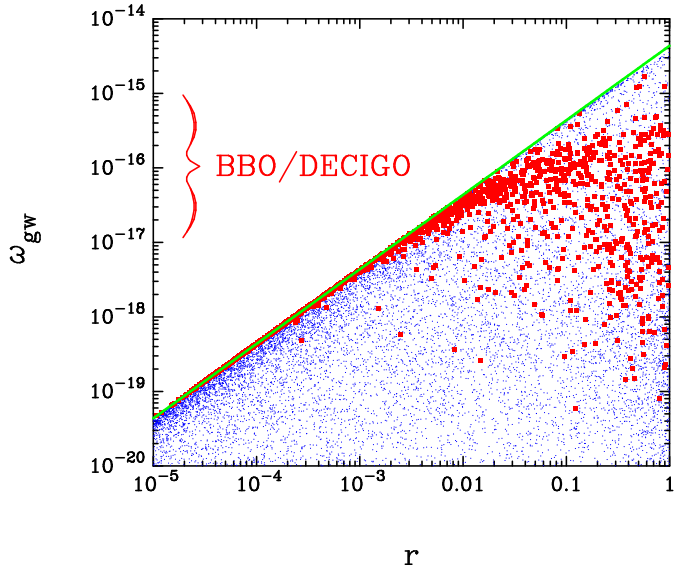}

\caption{Plots of gravitational wave spectrum $\omega_{\rm gw}$ against
tensor-scalar ratio $r$ for a large number of models evolved with
the inflationary flow equations. Square (red) points indicate models
satisfying the observational constraints on $n_s$ and $dn_s/d\ln k$
given by (\ref{niceobs2}).  The solid line shows the bound given by
Equation (\ref{upper}).}
\label{Figure5}
\end{figure}

Evaluating equation (\ref{gw1}) at wavenumbers characteristic of
space-based gravitational wave interferometers ($k \sim 6 \times 10^{13}\; {\rm
Mpc}^{-1}$, corresponding to frequencies $f \sim 0.1\; {\rm Hz}$) requires
a large extrapolation of the tensor power spectrum of about $16$ orders of
magnitude in scale from  CMB scales ($k_0 \sim 0.002 \;{\rm
Mpc}^{-1}$). Linking predictions for direct gravitational wave
detection to CMB polarization constraints is therefore model
dependent. Figure \ref{Figure5} shows results from\cite{chongchitnan}
based on the `inflationary flow' approach\cite{hoffman}\cite{easther}
in which the Hubble constant is parameterised as a polynomial,
\begin{equation}
H(\phi) = H_0 \left [1 + a_1\phi + a_2 \phi^2 + ..... + a_{n+1}
\phi^{n+1} \right ], \label{flow} 
\end{equation} 
truncated at finite $n$ (in our case, $n=10$) with the coefficients 
$a_1$,...$a_{n+1}$, drawn at random from some assumed  distribution
(see \cite{chongchitnan} for details).
The figure shows $\omega_{\rm gw} \equiv
\Omega_{\rm gw} h^{2}$ (where the Hubble constant, $H_0=100h \;{\rm
km}\;{\rm s}^{-1}{\rm Mpc}^{-1}$) evaluated at $N=20$ e-folds from the
end of inflation, corresponding to the time when perturbations with scales
relevant to direct detection experiments were equal to the Hubble
radius. The tensor-scalar ratio $r$ plotted in Figure \ref{Figure5} is
equation (\ref{tens1}) evaluated $N=60$ e-folds from the end of
inflation, corresponding to CMB scales $\sim k_0$.

 The (red) square points in Figure \ref{Figure5} show the subset of
models that satisfy the $2\sigma$ observational constraints
\cite{seljak}\cite{matteo} on $n_s$ and $dn_s/d\ln k$,
\begin{equation}
0.92\lesssim n_s \simlt  1.06, \quad -1.04\simlt dn_s/d\ln k \simlt 0.03,
\label{niceobs2}
\end{equation}
and the (green) line shows the bound derived by assuming that the Hubble
parameter $H(\phi)$ remains constant between $N=60$ and $N=20$:
\begin{equation}
\omega_{\rm gw} \approx 4.36\times10^{-15} r. \label{upper}
\end{equation}
Also shown is the rough sensitivity range\cite{kudoh}\cite{corbin} for
the proposed `post-LISA' space-based interferometers BBO and
DECIGO\footnote{Laser Interferometer Space Antenna (LISA); Big-Bang
Observer (BBO); Deci-Hertz Interferometer Gravitational-wave Observer
(DECIGO).}. The sensivities of these proposed missions are highly
uncertain and depend on the precise experimental configuration and
useable bandwidth (for example, unresolved white-dwarf binaries could
dominate the signal at frequencies below $\sim 0.2\; {\rm Hz}$
significantly reducing the sensitivities). The sensitivity range shown
in Figure \ref{Figure5} is meant to be indicative only.

\begin{figure}

\vskip 2.6 truein

\includegraphics{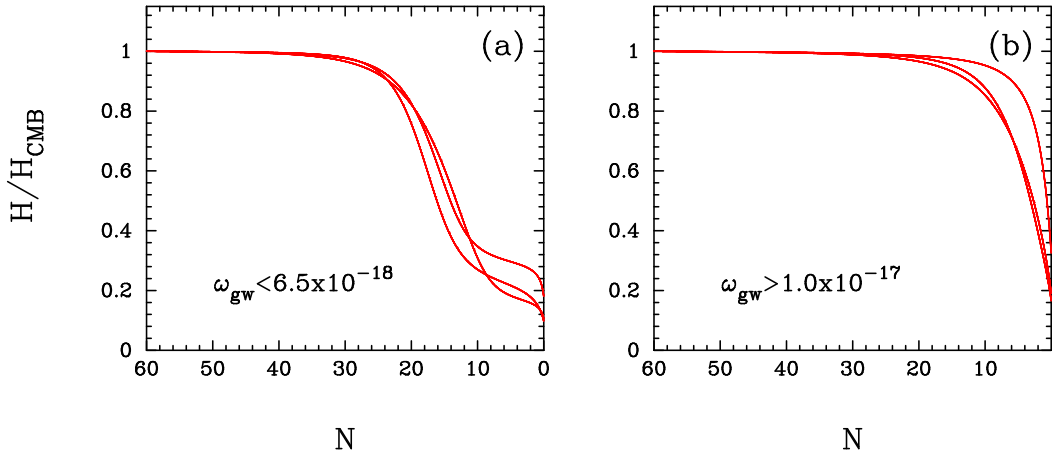}

\caption{Trajectories $H(N)$, from CMB scales ($N\approx 60$) to the
end of inflation $(N=0)$, for models with low tensor-scalar ratios at
CMB scales ($r$ in the range $2 \times 10^{-3}$ to $3 \times 10^{-3}$)
evolved using the inflationary flow equations.  The models plotted in
panel (a) have gravitational wave amplitudes at direct detection
scales of $\omega_{\rm gw} < 6.5 \times 10^{-18}$, whilst those shown
in panel (b) have higher amplitudes $\omega_{\rm gw} > 1 \times
10^{-17}$.  All of these models satisfy the observational constraints
on $n_s$ and $dn_s/d\ln k$ given by Equation (\ref{niceobs2}).}
\label{Figure6}
\end{figure}

Evidently, equation (\ref{upper}) provides a strict upper bound to the
gravitational wave spectrum independent of the shape of the inflationary
potential. However, the majority of models satisfying the
observational constraints (\ref{niceobs2}) give $\omega_{\rm gw}
\simlt 3 \times 10^{-16}$. It is difficult to exceed this value, even
if the tensor-scalar ratio is high at CMB scales, unless the shape
of the potential is adjusted to produce a sharp decline in $H(\phi)$
within the last $20$ e-folds of inflation. For $r \ll 1$ most models
lie very close to the limit (\ref{upper}) and even more fine-tuning of
the inflaton potential shape is required to generate models that lie
well below this bound (see Figure \ref{Figure6}). Notice that the
models shown in Figure \ref{Figure6}b have a sharp downturn in
$H(\phi)$ at the very end of inflation, it can be argued that {\it
all} models with $r \ll 1$ require fine-tuning. We will return to this
point in the next Section (which contains some cautionary remarks on
the concept of fine-tuning applied to inflationary models).

Adopting an optimistic sensitivity for a BBO/DECIGO mission of
$\omega_{\rm gw} \sim 10^{-17}$, Figure \ref{Figure5} shows that a
detection of gravitational waves from inflation is not expected unless
$r \simgt 0.002$ on CMB scales. This is not too different from the
sensitivity that seems feasible from the ground based CMB experiments
(Section 3.2), on a much shorter timescale and at a tiny fraction of
the cost\footnote{For a summary of other physical mechanisms that
could give rise to a cosmological background of gravitational waves,
{see\cite{chongchitnan}}.}.  There has been some discussion in the
literature of an `ultimate' DECIGO {interferometer\cite{kudoh}}, with a
sensitivity of $\omega_{\rm gw} \sim 10^{-20}$ limited by quantum
noise for $100 \;{\rm kg}$ test masses. Such a sensitivity, if it could
be achieved, would probe models with $r \simgt 10^{-6}$, far below the
levels likely to be reached with the CMB. Is it worth investing huge
resources to reach this type of limit? This is discussed in the next Section.

\section{Theoretical Implications}

In this Section, we will address the two questions raised in the Introduction.

\subsection{What does inflation predict for the amplitude of the tensor fluctuations?}

It is well known that the amplitude of the tensor mode CMB anisotropy
fixes the energy scale of inflation\cite{lyth}
\begin{equation}
V^{1/4} \approx 3.3 \times 10^{16} r^{1/4} \; {\rm GeV}. \label{theory1}
\end{equation}
The current upper limit of $r \simlt 0.36$ gives the constraint
$V^{1/4} \simlt 2.6 \times 10^{16}\; {\rm GeV}$, or equivalently $V
\simlt 2.2 \times 10^{-11} \M^4$.  {\it At present, there are no
compelling theoretical arguments to favour any particular energy
scale.} Since this energy scale depends only on the quarter power of
$r$, our experimental colleagues have to work extremely hard to
tighten the bounds:
\medskip
\begin{center}
\begin{tabular}{lcl}
Experiment    &  tensor-scalar limit  &    $V^{1/4}$ (GeV) \\
Planck & $r \sim 0.1$ &    $1.8 \times 10^{16}$  \\
Clover/QUIET & $\;\;r \sim 0.01$ &    $1.0 \times 10^{16}$  \\
BBO/DECIGO & $\;\;r \sim 10^{-3}$ & $5.9 \times 10^{15}$ \\
ultimate DECIGO & $\;\;r \sim 10^{-6}$ & $1.0 \times 10^{15}$
\end{tabular}
\end{center}
\medskip
\medskip
\noindent
Yet, given what we know about fundamental physics, the energy scale of
inflation could easily be $\sim 10^{14}\; {\rm GeV}$ or less, giving $r \simlt
10^{-10}$ which is well below the limit of any conceivable experiment.

 What then, should we make of the wide spectrum of opinion amongst
theorists, some of whom\cite{linde}\cite{steinhardt}\cite{boyle1}
argue that the tensor-scalar ratio should be measurably high and
others\cite{liddle} who argue that it should be immeasurably small?
We will review some of the arguments:

\medskip

\noindent
(i) {\it Initial Conditions:} Let us imagine that inflation begins in a patch
of about the Planck size, at Planck energies, and with an entropy (in
Planck units) $S \sim 1$. We know empirically (equation \ref{theory1})
that the classical fluctuations that we see in the Universe today 
were frozen at
much lower energy. So let us `connect' the Planck scale to this lower 
energy scale by assuming simple power-law potential:
\begin{equation}
V(\phi) =   \lambda \M^4 \left ( {\phi \over \M} \right )^\alpha. \label{theory2}
\end{equation}
Inflation occurs for field values, $\alpha/(4 \pi)^{1/2} \simlt
\phi/\M \simlt \lambda^{-1/\alpha}$, and the amplitude of the scalar
fluctuations in our Universe can be explained for suitably small
values of the parameter $\lambda$. (For example, for the quartic
potential, we require $\lambda \sim 4 \times 10^{-14}$). Notice that
the field values vastly exceed the Planck scale initially (though the energy
density is, by construction, always less than the Planck scale). In
fact, the field value at the time that the fluctuations on CMB scales
were equal to the Hubble radius is $\sim (N \alpha/4\pi)^{1/2}\M$
(at $N$ e-folds before the end of inflation).
The tensor-scalar ratio and scalar spectral index  in such models are,
\begin{equation}
r \approx  {4 \alpha \over N} \approx {\alpha \over 15}, \quad n_s \approx 1 - {2 + \alpha \over 2 N}
\approx 1 - {2 + \alpha \over 120}, \label{theory3}
\end{equation}
where we have assumed $N \approx 60$. It is worth noting that there is
already observational pressure on this type of model. The quartic
potential, $\alpha \approx 4$, is marginally excluded by
observations\cite{seljak} (mainly because of the large tilt in $n_s$)
but a quadratic potential provides an acceptable fit to the data.

The {\it initial conditions} provide the main motivation for this
class of models, since the Planck scale is linked to the much lower
energy scale at which the fluctuations on CMB scales were frozen.  For
potentials with $\alpha$ of order unity, the tensor amplitude must 
necessarily be high. There is, therefore, a very good
prospect of excluding this class of model with the next generation of
CMB polarization experiments.

\medskip

\noindent
(ii) {\it Fine-Tuning:} Another set of arguments that have been used to favour
inflationary models with a high tensor amplitude is based on 
fine-{tuning\cite{steinhardt}\cite{boyle1}}. 
 In a simple single field inflationary model, inflation
ends when the `slow-roll' parameter $\epsilon = 1$, yet the
fluctuations on CMB scales were frozen $N \sim 60$ e-folds from the
end of inflation when the field was rolling slowly. This introduces
`natural' values for the gradients of the inflaton
potential. Steinhardt\cite{steinhardt} argues that the most natural
values for the gradients are
\begin{equation}
 {V^\prime \over V} \sim {V^{\prime\prime} \over V^\prime} \sim {1 \over N}, \label{theory4}
\end{equation}
leading to the expectations,
\begin{equation}
r \sim  {14 \over N}, \quad n_s \sim 1 - {3 \over  N}, \label{theory5}
\end{equation}
similar to the chaotic inflation values of equation
(\ref{theory3}). An attempt to quantify the degree of fine-tuning
required to violate (\ref{theory5}) involves counting the zeros that
$\epsilon$ or $\eta$ undergo during the last 60 e-folds of
{inflation\cite{boyle2}}. An impression of the fine-tuning involved is
given by panel (b) of Figure \ref{Figure6}. All of the models shown
here, with a tensor-scalar ratio $r \sim 10^{-3}$, show a sharp
downturn in $H(\phi)$ within the last few e-foldings of inflation.

\begin{figure}

\vskip 3.2 truein

\includegraphics{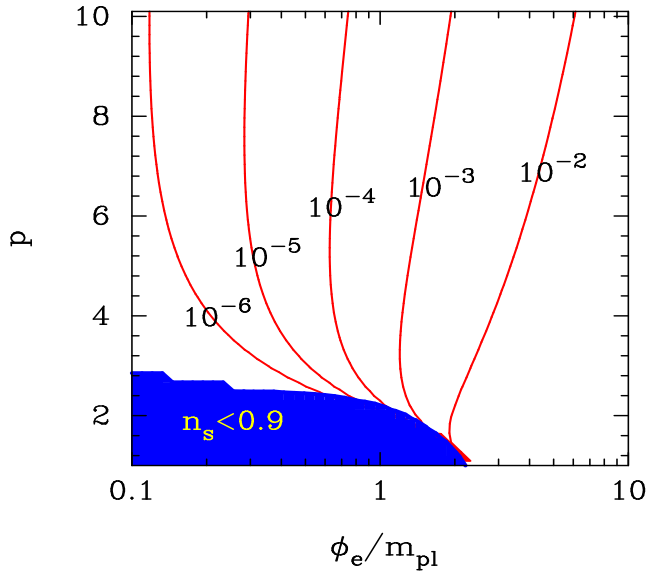}

\caption{Contours of the tensor-scalar ratio $r$ for inflation with
the potential (\ref{potential}) (starting from $\phi \approx 0$) as a
function of the parameters $p$ and $\phi_e$. Models with parameters 
in the shaded (blue) region have an unacceptably red scalar spectral index
$n_s < 0.90$, or do not inflate by $60$ e-folds.}
\label{Figure7}
\end{figure}

How impressed should we be with this type of fine-tuning argument?
Consider the potential\cite{kinney2}\footnote {This potential violates
condition (4) of Boyle and collaborators{\cite{boyle1}}, in that it
does not evolve smoothly to an analytic minimum with $V \approx
0$. This condition seems overly restrictive to us since complex physics
is inevitable as $V(\phi)$ plummets from the energy scale of inflation to
the present vacuum energy scale of a few milli-eV.}
\begin{equation}
V(\phi) = V_0 \left( 1 - \left ( { \phi \over \phi_e} \right )^p \right ). \label{potential}
\end{equation}
We have solved the Hamilton-Jacobi equation,
\begin{equation}
 \left(H^\prime(\phi)\right)^2-{12\pi\over{\M^2}}H^2(\phi)=
-{32\pi^2\over\M^4}V^\prime(\phi)\label{ham}
\end{equation}
starting from $\phi \approx 0$, evaluating `observables' ($n_s$, $r$
{\it etc.}) $60$ e-folds from the end of inflation (when the field
$\phi$ is always in the slow-roll regime).  Figure 7 shows contours of
the tensor-scalar ratio as a function of the parameters $p$ and
$\phi_e$. Models in the shaded region either do not inflate for $N =
60$ e-folds, or produce a scalar spectral index with an unacceptably
red tilt of $n_s < 0.9$. Evidently, the tensor-scalar ratio is largely
controlled by the parameter $\phi_e$, which can be adjusted to produce
tensor-scalar ratios that are unobservably small. It might be argued
that for $p>2$ the absence of a $\phi^2$ mass term at small field
values  requires some form of
fine-tuning or special symmetry, but is the shape of the potential
unreasonably contrived?  We let the reader judge.

Another fine-tuning argument goes as {follows\cite{steinhardt}}: The
amplitude of the scalar fluctuation spectrum requires an energy scale
of inflation of
\begin{equation}
V^{1/4} \sim 10^{-5/2} (1 + w)^{1/4}\M,  \label{s1}
\end{equation}
where $w$ is the equation of state parameter $w = p/\rho$. Now $1+w$
must necessarily be much smaller than unity during the inflationary
phase, but special fine-tuning is required to give $V^{1/4} \ll \M$.
For example, $V^{1/4} \sim 10^{-6} \M$ requires $(1+w) \sim 10^{-14}$.
But in slow-roll inflation, the equation of state parameter is related
to the gradient of the potential:
\begin{equation}
 (1 + w) \approx {2 \over 3} \epsilon_V, \quad \epsilon_V \equiv {\M^2 \over 16\pi}
\left( {V^\prime \over V} \right)^2 , \label{s2}
\end{equation}
and the number of e-folds of inflation is given by,
\begin{equation}
N(\phi) \approx {8 \pi \over \M^2} \int { V \over V^\prime}\;d\phi =
{2 \sqrt{\pi} \over \M} \int {d\phi \over \sqrt{\epsilon_V(\phi)}}.
\label{s3}
\end{equation}
Models with very small $\epsilon_V$ will therefore inflate by many
e-folds, and so weighting by volume will strongly favour models with
very flat potentials. But should we weight by volume? This is an
example of the {\it measure} problem that plagues cosmology (see, for
example, the papers by Tegmark\cite{tegmark} and
{Vilenkin\cite{vilenkin}}, and references therein). Until we have a
more complete understanding of the measure problem, it seems dangerous
to place much weight on fine-tuning arguments such as those discussed
here.

\medskip

\noindent
(iii) {\it Agnosticism:} Chaotic inflation is an example of what are
sometimes called `high-field' models of {inflation\cite{kinney}},
since field values must necessarily exceed the Planck scale.  It has
been argued\cite{lyth2} that this is an unattractive feature, since
quantum gravity corrections would render an effective field theory out
of control at $\phi \simgt \M$. However, as
Linde\cite{linde1}\cite{linde2} has emphasised, quantum gravity
corrections to $V(\phi)$ should become large only for $V(\phi) \simgt
\M^4$. (Linde also discusses some phenomenological models with high field values
in the context of supergravity, in which the
potential displays a shift-symmetry $\phi \rightarrow \phi + {\rm
constant}$). Although the fundamental physics behind high-field
inflationary models is poorly understood, it is premature to exclude
them at this stage.

Inflationary models with small field values $\phi \ll \M$ {\it always}
produce a negligible tensor {amplitude\cite{lyth2}\cite{efstathiou}}.
Thus, some authors who approach inflationary model building from a more
`traditional' particle-physics perspective \cite{liddle}\cite{lyth3}
argue that the tensor-scalar ratio should be negligibly small. But
this type of argument is unpersuasive because, as we have stressed
above, insisting on a `controllable' effective field-theory may just
reflect our lack of knowledge of fundamental physics.

In our view, the prudent position at this stage is agnosticism. We
simply do not know whether inflation is high-field, or low-field,
high-energy or low-energy. For example, in the influential
`string-inspired' brane-inflation construction of Kachru and
collaborators\cite{kachru}, 
(the \KKLMMT scenario) the specific example given has parameters:
\begin{equation}
\left. \begin{array}{c}
    V^{1/4} \sim 10^{14} \;{\rm GeV}, \\
    r \sim 10^{-10}, \\
    n_s \sim 0.97, 
\end{array} \right \}  \label{kachru}
\end{equation}
and so produces a negligible tensor amplitude. Of course, the \KKLMMT
model is speculative, but it illustrates that there is no guarantee that
the tensor modes will lie within a range  accessible to  experiment.

\subsection{Is it really worth building experiments that can only cover
a small range of tensor amplitudes?}

In the previous Sections we have argued that it is feasible to design
experiments (at relativeley low cost) to probe a tensor-scalar ratio of
$r \sim 10^{-2}$. A failure to detect tensor modes from inflation at
this level would rule out the chaotic inflationary models described
above and other examples of `high-field' inflation. This is a well
motivated and achievable goal.

 But if we fail to detect tensor modes at this level, what then? Do we
continue the search with a next-generation `CMBpol' satellite designed
to reach\footnote{Assuming that the Galactic polarization and lensing
signals can be subtracted accurately at this level.} $r \sim
10^{-4}$?  The range of energy scales probed by such an
experiment is so miniscule that the case would seem weak {\it unless
there are strong theoretical reasons to favour this narrow
range}. There are no such reasons at present.

  However, a failure to detect tensor modes at a level of $\sim
10^{-2}$ surely points to flat potentials and to an abrupt end to
inflation.  It therefore seems more sensible to design experiments to
test for signatures associated with this abrupt end, rather than to
focus single-mindedly on a search for inflationary tensor modes that
will, in all likelihood, prove fruitless.

Producing an abrupt end to inflation is one of the main motivations
for hybrid inflationary {models\cite{liddle}}.  The archetypal hybrid
inflation model\cite{linde3} uses two coupled scalar fields with a
potential,
\begin{equation}
 V = V_0 + {1 \over 2} m_\phi^2 \phi^2 - {1 \over 2}
m^2_\psi \psi^2 + {1 \over 4} \lambda \psi^4 + {1 \over 2}
\lambda^\prime \psi^2 \phi^2. \label{hybrid1}
\end{equation}
In this model, most of the energy density during inflation is supplied
by the field $\psi$.  Once $\phi$ rolls down to a critical value
$\phi_{\rm c} = m_\psi/\sqrt{\lambda^\prime}$, the $\psi$ field is
destabilised and rolls down to its true vacuum, ending inflation.  In
some realisations of hybrid inflation heavy cosmic strings may be
formed during a phase transition at the end of inflation that could
produce a CMB anisotropy of comparable amplitude to the scalar
{fluctuations\cite{linde4}\cite{jeannerot}}. Cosmic strings could,
in principle, provide an observable signature of the end of inflation
even if the tensor-scalar ratio, $r$, is unobservably small.

More recently, the realisation that our Universe might be confined to
a brane has stimulated a lot of research on whether interactions
between branes can produce inflation. (See {Quevedo\cite{quevedo}},
for a review). These models (of which the \KKLMMT scenario is an
example) are speculative at present, but they share some features that
may be generic. Firstly, in these models the inflaton field is
identified with the separation between brane, or brane-anti-brane,
pairs within a 5-D `bulk'.  Thus inflation acquires a geometrical
interpretation. Secondly, open strings must end on brane pairs. Below
some critical separation, the mass of an open string mode becomes
negative (tachyonic) at which point inflation ends abruptly. This
mechanism for producing a sharp end to inflation is an attractive
feature of brane inflation. Thirdly, cosmic superstrings (or ordinary
gauge strings) can form at the end of inflation. The expected tensions
of cosmic superstrings are model dependent, but may plausibly lie in
the range\cite{pogosian}\cite{polchinski}\cite{polchinski1}
\begin{equation}
10^{-11} \simlt G \mu \simlt 10^{-6} . \label{string1}
\end{equation}
If the string tension lies towards the upper end of this range\footnote
{Note that constraints from pulsar timing may already overlap with this
upper {bound\cite{kaspi}}.}, then
strings should be easily detectable in the CMB, both as non-Gaussian
signatures in the temperature maps and through their distinctive
$B$-mode polarization power spectrum (which is very
different\cite{pogosian} from the $B$-mode spectrum expected from
tensor modes, peaking at $\ell \sim 1000$).

Recently Damour and Vilenkin\cite{damour}\cite{damour1} have
calculated the gravitational wave spectrum from bursts associated with
cusps and kinks in loops of cosmic superstrings as a function of the
theoretically uncertain intercommutation
probability\footnote{Fundamental strings differ from ordinary gauge
strings in their reconnection properties, since they can miss each
other in higher dimensions. The reconnection probability may therefore
be very much less than unity for fundamental {strings\cite{jackson}},
whereas it is very close to unity for gauge strings.}.  They conclude
the gravitational wave bursts from cosmic superstrings with tensions
as low as $G\mu\sim 10^{-14}$ should be detectable by LISA and may even be
observable by ground-based detecters such as LIGO if $G\mu \simgt
10^{-10}$ and the reconnection probability is small. As
Polchinski\cite{polchinski} has emphasised, cosmic superstrings could
be the brightest objects visible in gravitational wave astronomy.

In summary, CMB experiments designed to probe tensor-scalar ratios as
low as $r \sim 10^{-2}$ are feasible and well motivated. They could
rule out chaotic inflation and many other versions of `high-field'
inflation. If tensor modes are not detected at this limit, then this
suggests flat inflationary potentials and an abrupt end to inflation.
Designing an experiment to probe as low as $r \sim 10^{-4}$ is a
formidably daunting prospect, and yet this would improve the limit on
the energy scale of inflation by only a factor of $\sim 3$. This seems
poorly motivated, because there is no particular reason to expect the
energy scale of inflation to lie in the narrow range $3.3 \times
10^{15} \;{\rm GeV} \simlt V^{1/4} \simlt 1 \times 10^{16}\; {\rm GeV}$. The
energy scale of inflation could easily be $\sim 10^{14}\;{\rm GeV}$ 
or less.
It may be more profitable to search for signatures associated with an
abrupt end to inflation, such as non-Gaussianity in the CMB from
multi-field inflation,  cosmic strings formed at the end of inflation,
$B$-mode polarization in the CMB associated with cosmic strings, and
gravitational wave bursts from cosmic string cusps.

\section*{Acknowledgements}
This work is supported by the UK Particle Physics and Astronomy
Research Council. SC acknowledges the award of a Dorothy Hodgkin
studentship. We thank Anthony Challinor for contributing Figures 1-4.
We thank  Anthony Challinor and Antony Lewis for comments on the 
manuscript.

%

\end{document}